\documentclass[12pt,preprint,mathscr]{aastex}
\usepackage{euscript}

\newcommand{\hehp}{HeH$^+$}
\newcommand{\htp}{H$_3^+$}
\newcommand{\hh}{H$_2$}
\newcommand{\hehep}{He$_2^+$}
\newcommand{\hhm}{H$_2^-$}
\newcommand{\hhp}{H$_2^+$}
\newcommand{\etal}{{\it et al.}}
\newcommand{\MNRAS}{Mon. Not. R. Astron. Soc.}

\newcommand{\gcc}{g cm$^{-3}$}
\newcommand{\lnhnhe}{$\log_{10}(N_{\rm H}/N_{\rm He})$}
\newcommand{\AAA}{A\&A}
\newcommand{\teff}{$T_{eff}$}
\newcommand{\HII}{H~{\scshape II}}
\newcommand{\HeII}{He~{\scshape II}}
\newcommand{\HeI}{He~{\scshape I}}

\shorttitle{\hehp\  in white dwarfs.}
\shortauthors{Harris et al.}

\begin{document}

\title{The role of \hehp\  in cool helium rich white dwarfs.}

\author{G. J. Harris, A. E. Lynas-Gray\altaffilmark{1}, S. Miller and J. Tennyson\altaffilmark{2}.}
\affil{Department of Physics and Astronomy, University College London, London, WC1E 6BT, UK.}


\altaffiltext{1}{Permanent address: Department of Physics, University of Oxford, Keble Road, Oxford OX1 3RH, UK.}        
\altaffiltext{2}{Corresponding Author:  j.tennyson@ucl.ac.uk}

\begin{abstract}
\hehp\  is found to be the dominant positive ion over a wide range of temperatures and densities relevant to helium rich white dwarfs. The inclusion of \hehp\  in ionization equilibrium computations increases the abundance of free electrons by a significant factor. For temperatures below 8000~K, He$^-$ free-free absorption is increased by up to a factor of 5, by the inclusion of \hehp. Illustrative model atmospheres and spectral energy distributions are computed, which show that \hehp\  has a strong effect upon the density and pressure structure of helium rich white dwarfs with \teff$<$8000~K. The inclusion of \hehp\  significantly reddens spectral energy distributions and broad band color indices for models with \teff$<$5500~K. This has serious implications for existing model atmospheres, synthetic spectra and cooling curves for helium rich white dwarfs.
\end{abstract}


\section{Introduction}

\citet{Bergeron02} analyzed the recently discovered white dwarfs SDSS~J133739 +000142 and LHS~3250 \citep{HHarris99,HHarris01}, identifying both objects as extreme helium rich cool white dwarfs. However they encountered significant problems when attempting to fit the spectral energy distributions (SEDs). \citet{Bergeron02} concluded that the discrepancy between their SEDs and the observed fluxes, is due to the physics used to calculate their model atmospheres. Here we investigate the molecular ion \hehp\  as part of the missing physics of helium rich white dwarfs. We demonstrate that the opacity of a helium rich white dwarfs photosphere is significantly affected by \hehp. From the discussion of \citet{Fontaine}, it follows that increased opacity arising from \hehp, will lengthen the cooling time for helium rich white dwarfs with \teff$<$ 8000~K.

The only attempts to study \hehp\  in helium rich white dwarfs known to us was made by \citet{Gaur88,Gaur91}. They showed that \hehp\  exists in significant quantities in helium rich white dwarfs and suggested a search for the infrared lines of \hehp.


\section{Equation of state}
\label{sec:EoS}

The equation of state (EoS) is a vital component of any model atmosphere, it links the state parameters such as temperature, pressure, density, and internal energy. It also calculates the relative abundance of each species within the gas, which are essential to obtain accurate radiative opacities. The photospheres of cool extremely helium rich white dwarfs have densities which can reach upward of 1 \gcc, under such conditions the use of a non-ideal EoS is required.

We have adapted the non-ideal H/He EoS of \citet{Luo}. This EoS accounts for the non-ideal effects of electron degeneracy, Coulomb coupling and pressure ionization, but lacks an accurate treatment of pressure dissociation. The abundance of \hh\  is estimated using an equilibrium constant for the reaction: H$_2 \rightleftharpoons$ 2H, so that \hh\  pressure dissociates as hydrogen pressure ionizes.

To account for the pressure ionization of H$^-$ we have added a term to the hydrogen ionization equilibrium, given by eq. (22) and (23) in \citet{Luo}, so that
\begin{eqnarray}
y_{H^-} & = & L_{H^-}/L_{H} \nonumber \\
L_H & = & L_{H I} + L_{H {\small II}} + L_{H^-} \label{eq:ionfrac}
\end{eqnarray}
where $y_{H^-}$ is the ionization fraction of atomic and ionic hydrogen in the form of H$^-$, $L_{\rm H I}$ and $L_{\rm H II}$ are the grand partition functions of atomic hydrogen and a proton (see \citet{Luo} eq. 23). The grand partition function of H$^-$ is given by
 \begin{equation}
L_{H^-} = W_{H^-} \exp(2\lambda-E_{H^-}/kT) \label{eq:grandz}
\end{equation}
where $\lambda$ is the electron degeneracy, $E_{H^-}$ is the sum of the ionization potential of hydrogen and H$^-$ (14.352 eV), and $W_{H^-}$ is given by eq. (11)-(16) in \citet{Luo} using a characteristic radius for H$^-$ of 1.15 \AA\  \citep{LenzuniSaumon}.

Under certain conditions, the trace ionic molecules \hhm, \hhp, \htp, \hehp, and \hehep\  are responsible for nearly all the free electrons in a H/He gas. We calculate equilibrium constants for the formation of \hh, \hhm, \hhp, \htp, \hehp, and \hehep\ from atomic H and He, H$^-$, and free electrons with the Saha equation. Subject to conservation of charge, and of H and He nuclei, the equilibrium constants and ionization fractions are used to construct 3 non-linear simultaneous equations. These 3 equations are solved using a multi-variable Newton-Raphson technique. In this way the number densities for each species can be calculated for any given temperature, pressure, hydrogen fraction and value of $\lambda$. The internal partition functions we use are detailed in \citet{Harris04}, for \hehp\  we use the partition function of \citet{Engel}. A converged value of $\lambda$ is found by iterating over a further conservation of charge equation:
\begin{eqnarray}
C_eT^{3/2}F_{1/2}(\lambda-\epsilon_{CC}/kT) & = & N_{{\rm H {\small II}}} + N_{{\rm He {\small II}}} + 2N_{{\rm He {\small III}}} - N_{\rm H^-} + N_{\rm H_2^+} - N_{\rm H_2^-}  \nonumber \\ 
 & & +N_{\rm H_3^+}+ N_{\rm HeH^+} + N_{\rm He_2^+} \label{eq:coneq}
\end{eqnarray}
where N$_x$ is the number density of species $x$, $\epsilon_{CC}$ is the free electron Coulomb coupling energy \citep{Luo}, $F_{1/2}$ is a Fermi-Dirac integral, T is temperature and a constant $C_e=(2^{1/2}/\pi^2)(km_e/\hbar^2)^{3/2}$. The left hand side of eq \ref{eq:coneq} is the number density of free electrons (see \citet{Luo1994,Luo}) and the right hand side counts the charge on all ions.

\begin{figure}
\plottwo{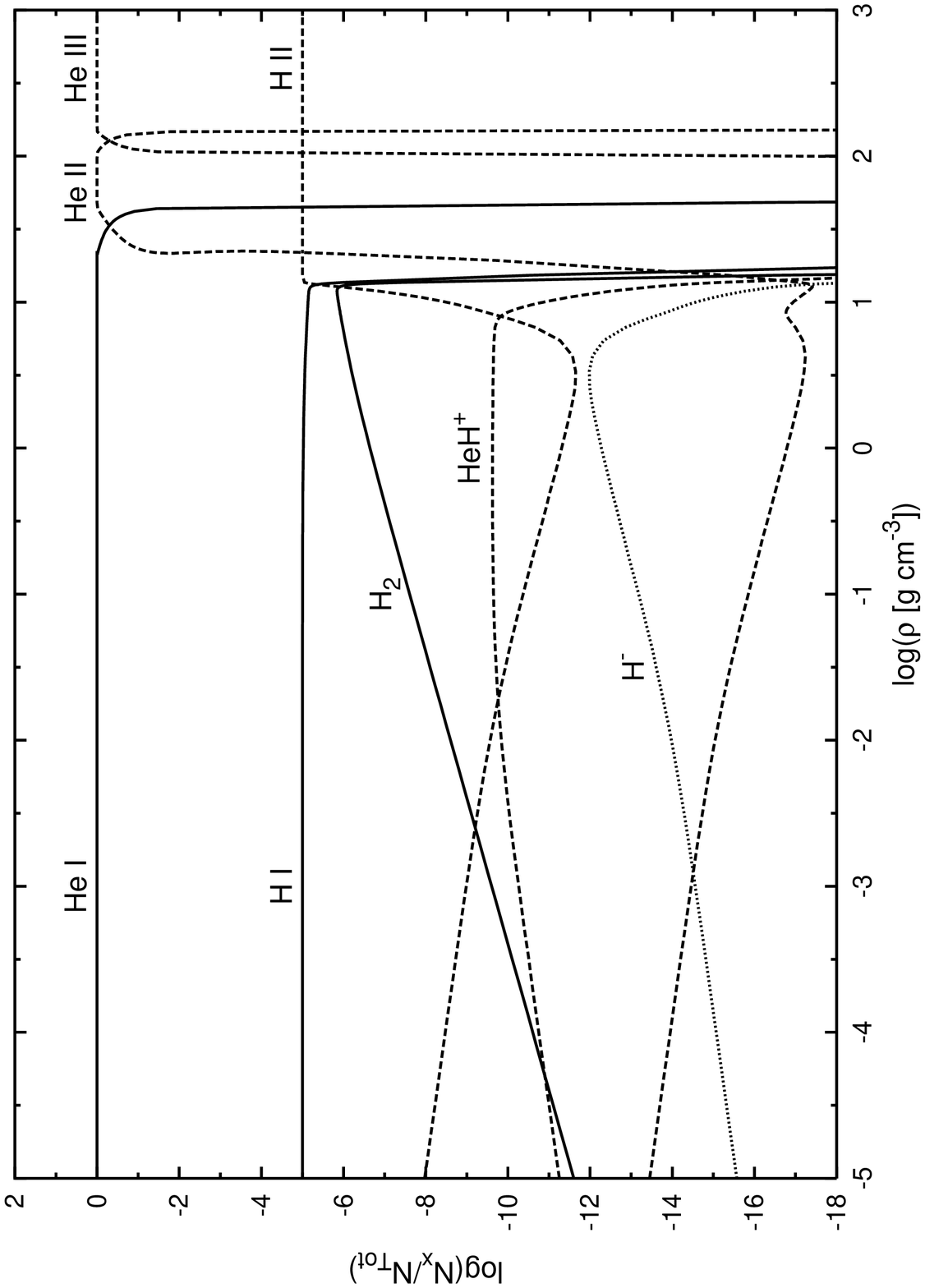}{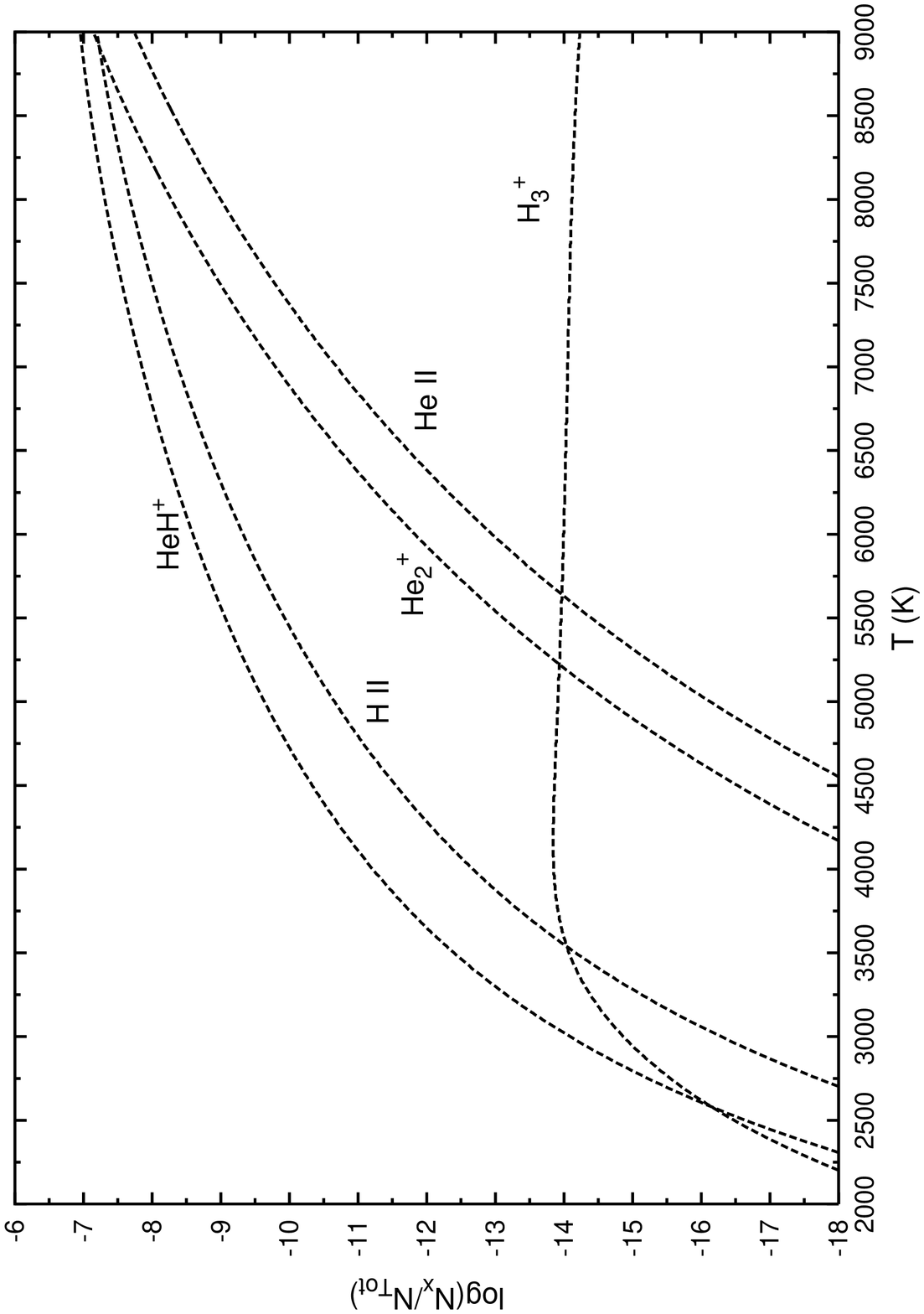}
\epsscale{.45}
\plotone{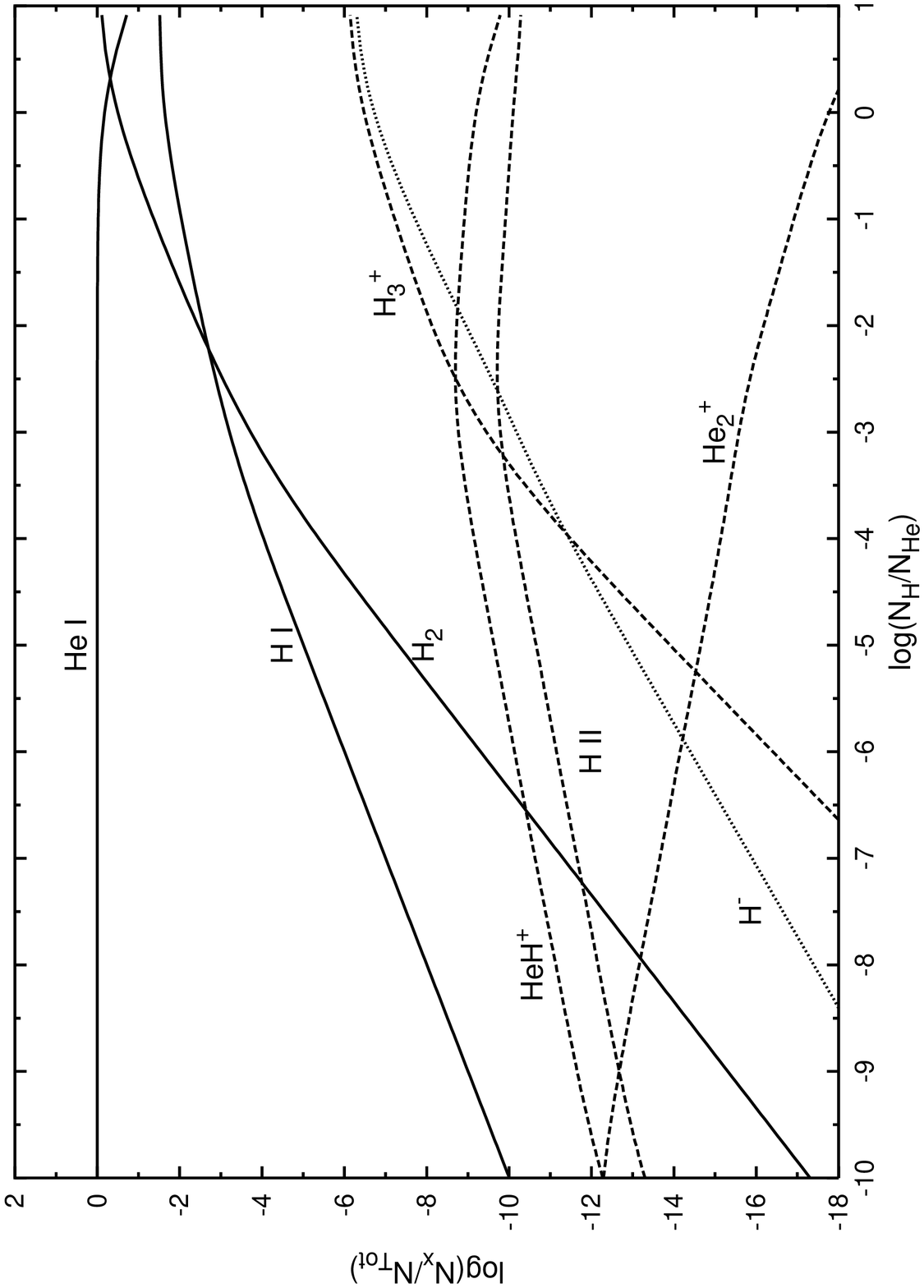}
\epsscale{1.0}
\caption{Chemical and ionization equilibrium as a function of \lnhnhe, density and temperature. Values of temperature of 5000~K, density of 0.2 \gcc, and \lnhnhe$=$--5 are used. Neutral species are given solid lines, positively charged species dashed lines, and negatively charged species dotted lines. \label{fig:eos}}
\end{figure}

Figure \ref{fig:eos} shows the number fraction of the species within our EoS, as a function of H to He number ratio, density, and temperature. 
At 5000~K and density of 0.2 \gcc\, \htp\  is the dominant positive ion for the hydrogen rich case. \hehp\  is the dominant positive ion for the helium rich range $-10 < \log_{10}(N_H/N_{He}) < -2.5$ and \hehep\  becomes the dominant positive ion for $\log_{10}(N_H/N_{He}) < -10$. Figure \ref{fig:eos} indicates that \hehp\  continues to be the dominant positive ion over a range of densities and temperatures. \citet{Lenzuni91} present an EoS and mean opacities for a H/He gas of 72\% hydrogen by mass, they correctly state that the opacity coefficent of \hehp\  is wholely irrelevant. However, as illustrated below, for a helium rich mix \hehp\  strongly affects the opacity and cannot be neglected.

\section{Opacity function}
\label{sec:opacity}

The opacity of a gas under the extreme pressures found in the photospheres of helium rich white dwarfs remains in question \citep{Iglesias,Bergeron02}. The opacity of a cool helium rich atmosphere is dominated by \hh-He collision induced and He$^-$ free-free absorption, and \HeI\  Rayleigh scattering \citep{Malo,Iglesias,Rohrmann}. As such the opacity is strongly dependent upon the abundance of free electrons and \hh. 
The sources of opacity data that we use is discussed in \citet{Harris04}. 

\begin{figure}
\epsscale{.45}
\plotone{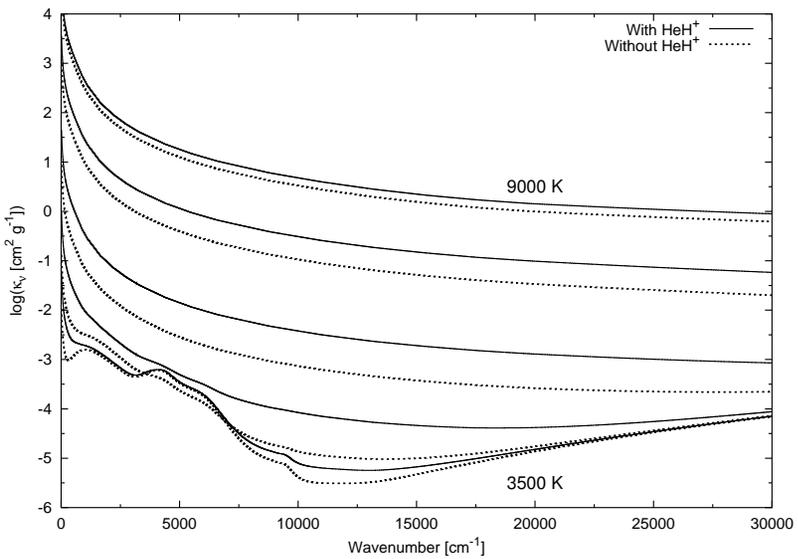}
\epsscale{1.0}
\caption{The continuous opacity function as a function of wavenumber at constant values of $\rho=0.5$ \gcc\  and \lnhnhe=--5, calculated at temperatures of 9000, 7000, 5000, 4000, and 3500~K. \label{fig:opa-t}}
\end{figure}

The monochromatic absorption coefficient at $\rho=0.5$~\gcc, \lnhnhe$=$--5, over a range of temperatures, computed both including and neglecting \hehp\  from our EoS, is shown in figure \ref{fig:opa-t}. It is evident that if \hehp\  is neglected the gas opacity can be underestimated by as much as a factor of 5, over a significant range of temperatures. The dominant opacity, across the frequency range shown in figure \ref{fig:opa-t} and for temperatures upward of 5000 K, is He$^-$ free-free absorption. At lower temperatures collision induced absorption, in the infrared, and \HeI\  Rayleigh scattering, in the visible/ultra-violet, become important and eventually take over from He$^-$ free-free. 

To determine if \hehp\  rotation-vibration lines would be observable in a helium rich white dwarf we have employed the recent publicly available \hehp\  linelist of \citet{Engel}. We find that the absorption lines of \hehp\  are too weak to overcome the continuous opacity, under the temperatures and densities found in helium rich white dwarfs.
Therefore \hehp\  lines will not be visible  in the spectra of helium rich white dwarfs. For a discussion of \hehp\  line opacity and some of the temperatures densities in which it is important see \citet{Engel}.

\section{Model atmospheres \& spectral energy distributions.}
\label{sec:model}

We use the plane parallel model atmosphere code {\scshape marcs} \citep{Gustafsson}, modified for the new non-ideal EoS subroutines, discussed in section \ref{sec:EoS}, and the new continuous opacity subroutines, discussed in section \ref{sec:opacity}. The new EoS and opacity function subroutines are fast enough to be run in real time.

As discussed in \citet{Saumon94} and \citet{Bergeron}, in the optically thin regions, the unusual opacity function of a metal free H/He gas results in multiple roots in the equation of radiative equilibrium.
The high temperature solution to radiative equilibrium in the optically thin regions is preferentially found in our models. Such a solution is not physically realistic, rendering our models of \teff$\leq$5000~K below $\log \tau_R=-2$ unreliable. However, as this only occurs at very small optical depths the emergent flux is unaffected. 

We also experienced a problem with convergence of the convective flux at temperatures of 5000~K and below. The pressure-temperature gradient ($\nabla$) is very close to the adiabatic gradient ($\nabla_{ad}$), so that $(\nabla-\nabla_{ad})/\nabla \sim 10^{-3}$ in the convective zone. In the cool highly non-ideal regions, numerical noise in the value of $\nabla_{ad}$ calculated within our EoS is of this order, resulting in convergence problems with the convective flux. We have therefore not been able to obtain converged models below \teff$=$4500~K.

\begin{figure}
\plottwo{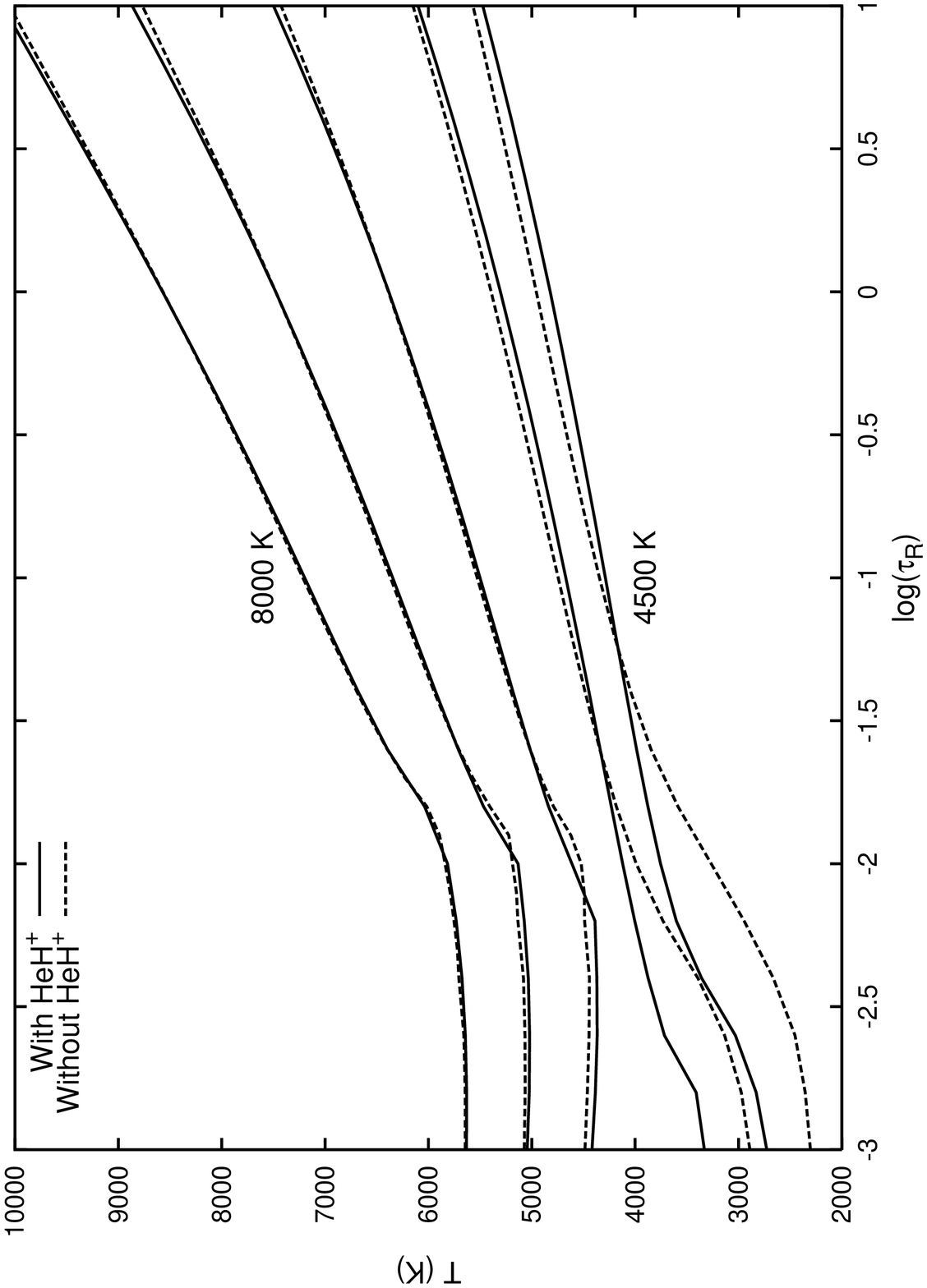}{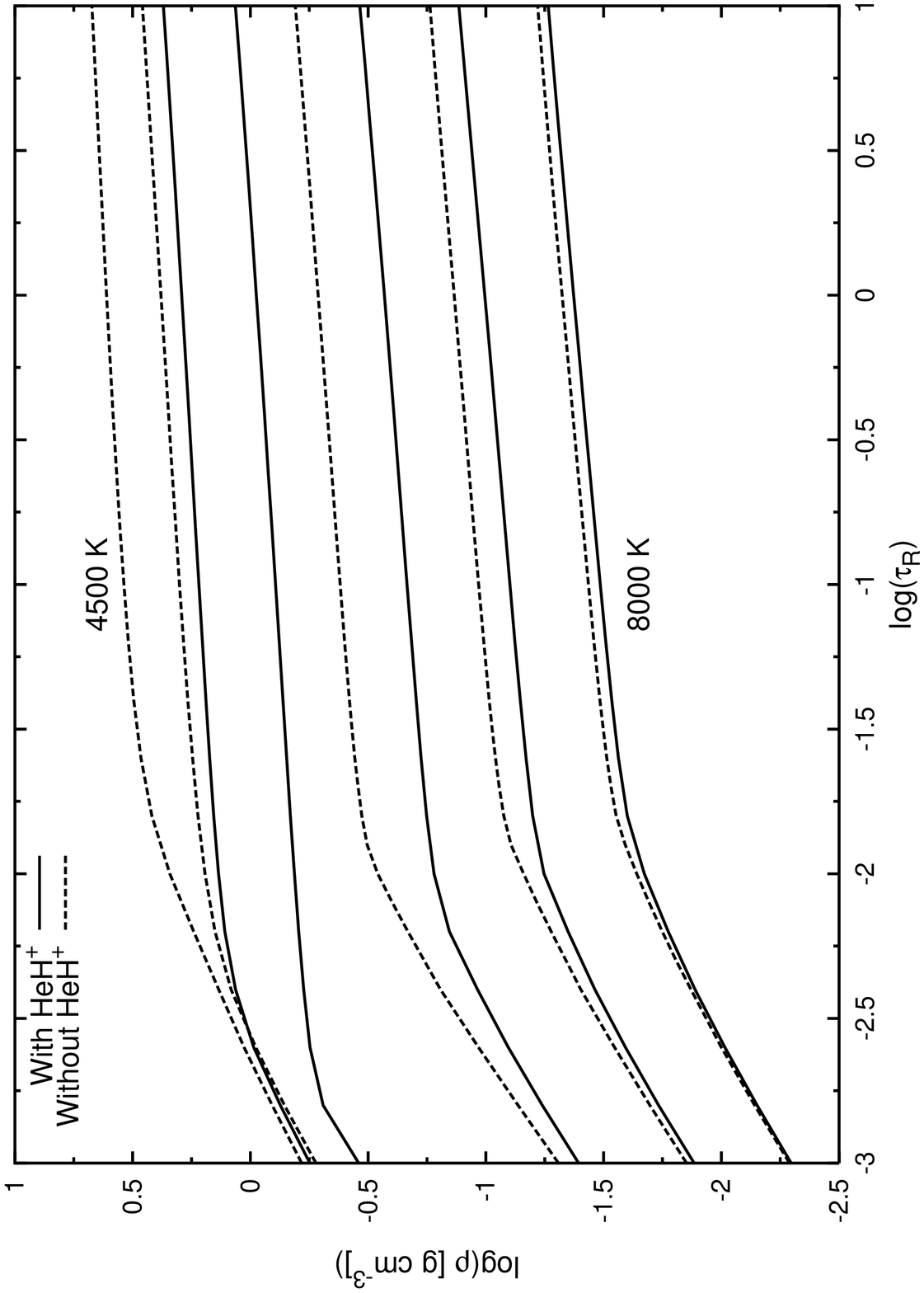}
\caption{Optical depth verses temperature and density for models of 4500, 5000, 6000, 7000, and 8000~K, computed including and neglecting \hehp\  in the ionization equilibrium, with \lnhnhe $=-5$.\label{fig:atm}}
\end{figure}

We have computed a set of model atmospheres for $\log g = 8$, \lnhnhe$=10^{-5}$, and between effective temperatures of 4500 and 8000~K, including and neglecting \hehp. Figure \ref{fig:atm} shows optical depth verses temperature and density for model atmospheres of 4500, 5000, 6000, 7000 and 8000~K. Although the temperatures remain relatively unperturbed by the inclusion of \hehp, there is a very strong affect upon the density and pressure. If \hehp\  is neglected then the density and pressure can be overestimated by up to a factor 5, similarly the electron pressure significantly underestimated. For \teff\  of $\geq$ 8000~K there are significant electrons released from \HII\  and \HeII, which reduces the importance of \hehp. 

\begin{figure}
\epsscale{.45}
\plotone{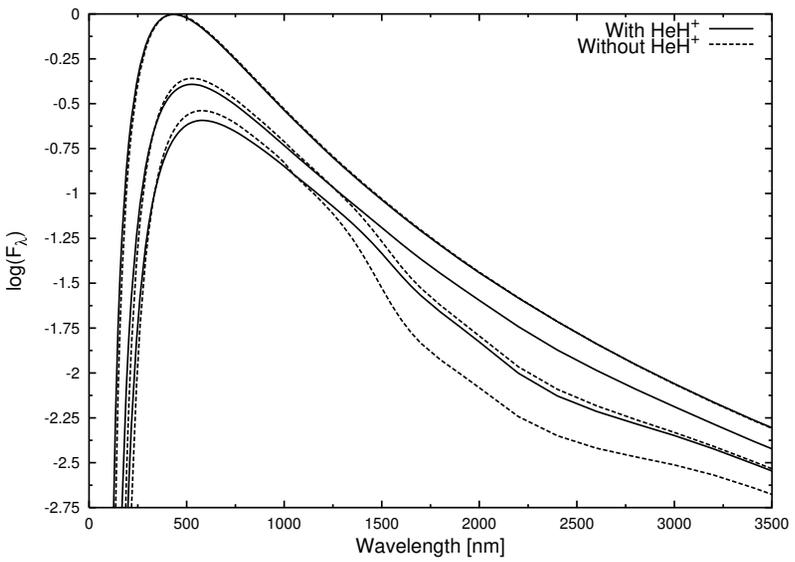}
\epsscale{1.0}
\caption{Spectral energy distributions for models of \teff$=$4500, 5000, and 6000~K. The logarithm of the relative flux is given per unit wavelength interval.\label{fig:SEDS}}
\end{figure}

Figure \ref{fig:SEDS} shows the SEDs of our 4500, 5000 and 6000~K models, with and without \hehp. The 4500 and 5000~K SEDs show a significant changes if \hehp\  is included in the ionization equilibrium, but the effect is only small for the 6000 K model. The reason for this is that above $\sim$5000~K He$^-$ free-free is the only significant source of opacity, so although the total opacity is increased the shape of the absorption function and hence SED is unchanged. For temperatures below 5500~K, He Rayleigh scattering and He-H$_2$ collision induced absorption contribute to opacity. As these opacity sources are unaffected by the increased abundance of electrons from \hehp, the increase in He$^-$ free-free opacity changes the shape of the total opacity function and SED. These differences are reflected in the broadband color indexes given in tables \ref{tab:1} and \ref{tab:2}. These colors were computed by using the bandpasses given by \citet{Bessell88,Bessell90} and calibrating using a spectrum of Vega. There are significant differences, at \teff$=$5500~K and below, between colors computed whilst including and neglecting \hehp. The large increase in the V--K magnitude, and most of the other color indices indicates that the models calculated with \hehp\  are significantly redder than the models calculated without \hehp, this is also apparent in the SEDs. In general all our colors are redder than the colors of \citet{Bergeron02}. 

\section{Conclusion}
\label{sec:conclusion}

A non-ideal H/He equation of state (EoS) which includes the molecular ion \hehp\  within the ionization equilibrium, has been presented. It has been demonstrated, that under helium rich conditions and over a range of temperatures and densities relevant to helium rich white dwarfs, \hehp\  is the dominant positive ion. Using the EoS, we have computed a set of continuous opacities which illustrate that \hehp\  can indirectly increase the opacity of a helium rich gas by up to a factor of 5. Using the recent \hehp\  linelist of \citet{Engel}, we have found that \hehp\  line opacity does not significantly contribute to the opacity at the densities found in helium rich white dwarfs. 

From a physical point of view one of the most interesting reasons for studying helium rich white dwarfs is that the densities of their photospheres access regions in which the gas is strongly non-ideal. \cite{Saumon91,Saumon95} have studied the pressure dissociation of \hh\  in a pure hydrogen environment. However, one of the shortcomings of our, and all other equations of state known to us is that there has been no study of the pressure dissociation of the important molecular ions, \htp. \hehp, and \hehep. Before we can fully understand helium rich white dwarfs, our understanding of the physics of cool dense H/He plasmas must be improved.

Our EoS and opacity function have been incorporated into a version of {\scshape marcs} \citep{Gustafsson}. Using this code we have computed model atmospheres, spectral energy distributions and broad band color indices for an illustrative range of helium rich white dwarfs. We find that in all models below 8000~K the pressure and density of the model atmospheres is reduced by up to a factor of five by the inclusion of \hehp. Furthermore, \hehp\ significantly reddens the SEDs and color indices, for models below \teff$=$ 5500~K. 
The importance of \hehp\  should prompt a review of all current model atmospheres, synthetic spectra and cooling curves for cool helium rich white dwarfs.

\acknowledgments

We thank Prof. Bengt Gustafsson for providing us with a version of {\scshape marcs}, Prof. Hugh Jones for providing a spectrum of Vega, and the UK Particle Physics and Astronomy Research Council (PPARC) for support.

\clearpage

\begin{table}
\begin{center}
\caption{Color indices for models calculated whilst neglecting \hehp. $\log g=8$ and \lnhnhe =$10^{-5}$. \label{tab:1}}
\begin{tabular}{rrrrrrrr}
\tableline\tableline
\multicolumn{1}{c}{\teff} &
\multicolumn{1}{c}{B--V} &  
\multicolumn{1}{c}{V--R} &   
\multicolumn{1}{c}{V--K} &    
\multicolumn{1}{c}{R--I} &    
\multicolumn{1}{c}{I--J} &    
\multicolumn{1}{c}{J--H} &    
\multicolumn{1}{c}{H--K} \\
\tableline
4500  & 0.85 & 0.52 & 0.58 & 0.48 & 0.30 & --0.54 & --0.17 \\
5000  & 0.72 & 0.44 & 0.81 & 0.42 & 0.35 & --0.18 & --0.22 \\
5500  & 0.60 & 0.38 & 1.14 & 0.36 & 0.32 & 0.12 & --0.04 \\
6000  & 0.50 & 0.32 & 0.98 & 0.30 & 0.24 & 0.12 & --0.01 \\
6500  & 0.42 & 0.27 & 0.77 & 0.25 & 0.18 & 0.09 & --0.03 \\
7000  & 0.36 & 0.23 & 0.59 & 0.21 & 0.12 & 0.07 & --0.05 \\
7500  & 0.30 & 0.20 & 0.43 & 0.17 & 0.07 & 0.05 & --0.06 \\
8000  & 0.25 & 0.17 & 0.30 & 0.14 & 0.03 & 0.03 & --0.08 \\
\end{tabular}
\end{center}
\end{table}

\begin{table}
\begin{center}
\caption{Color indices, for models calculated with \hehp. $\log g=8$ and \lnhnhe =$10^{-5}$. \label{tab:2}}
\begin{tabular}{rrrrrrrr}
\tableline\tableline
\multicolumn{1}{c}{\teff} &
\multicolumn{1}{c}{B--V} &  
\multicolumn{1}{c}{V--R} &   
\multicolumn{1}{c}{V--K} &    
\multicolumn{1}{c}{R--I} &    
\multicolumn{1}{c}{I--J} &    
\multicolumn{1}{c}{J--H} &    
\multicolumn{1}{c}{H--K} \\
\tableline
4500 & 0.84 & 0.52 & 1.32 & 0.51 & 0.49 & --0.03  & --0.18 \\
5000 & 0.70 & 0.44 & 1.44 & 0.43 & 0.42 & 0.16    & --0.01 \\
5500 & 0.59 & 0.37 & 1.23 & 0.36 & 0.32 & 0.16    & 0.02 \\
6000 & 0.50 & 0.32 & 0.98 & 0.30 & 0.24 & 0.13    & 0.00 \\
6500 & 0.42 & 0.27 & 0.80 & 0.25 & 0.18 & 0.10    & --0.03 \\
7000 & 0.35 & 0.23 & 0.59 & 0.21 & 0.12 & 0.07    & --0.04 \\
7500 & 0.30 & 0.20 & 0.43 & 0.17 & 0.07 & 0.05    & --0.06 \\
8000 & 0.25 & 0.17 & 0.29 & 0.14 & 0.03 & 0.03    & --0.08 \\
\end{tabular}
\end{center}
\end{table}

\clearpage


\end{document}